\documentclass[12pt,preprint,url]{aastex}
\usepackage{amsbsy}
\usepackage{amsmath}
\newcommand{\be}{\begin{equation}}
\newcommand{\ee}{\end{equation}}

\def\ltsima{$\; \buildrel < \over \sim \;$}

\def\lsim{\lower.5ex\hbox{\ltsima}}
\def\gtsima{$\; \buildrel > \over \sim \;$}
\def\gsim{\lower.5ex\hbox{\gtsima}}

\shorttitle{Fast fossil rotation of neutron star cores}
\shortauthors{A. Melatos}

\begin{document}
\title{Fast fossil rotation of neutron star cores}

\author{A. Melatos\altaffilmark{1}}

\email{amelatos@unimelb.edu.au}

\altaffiltext{1}{School of Physics, University of Melbourne,
Parkville, VIC 3010, Australia}

\begin{abstract}
\noindent 
It is argued that the superfluid core of a neutron star super-rotates
relative to the crust, because stratification prevents the core from
responding to the electromagnetic braking torque, until the relevant
dissipative (viscous or Eddington-Sweet) time-scale, which can exceed 
$\sim 10^3\,{\rm yr}$  and is much longer than the Ekman time-scale, 
has elapsed.
Hence, in some young pulsars, the rotation of the core today
is a fossil record of its rotation at birth,
provided that magnetic crust-core coupling is inhibited,
e.g.\ by buoyancy, field-line topology, or the presence of uncondensed
neutral components in the superfluid.
Persistent core super-rotation alters our picture of neutron stars in several ways, 
allowing for magnetic field generation by ongoing dynamo action
and enhanced gravitational wave emission from hydrodynamic instabilites.
\end{abstract}


\section{Introduction 
 \label{sec:fas1}}
Shear flows are endemic to neutron stars.
For example, the interpenetrating components of the stellar interior rotate differentially.
In the inner crust, an inviscid neutron condensate flows through a
rigid nuclear lattice.
Nuclear pinning controls the relative velocity 
\citep{has11,war11}.
The condensate also flows relative to a viscous fluid component, 
consisting of electrons, protons, and uncondensed neutrons.
Entrainment, hydromagnetic stresses, and Ekman pumping control 
the relative velocity; under certain conditions, the nuclear lattice
and viscous component lock together magnetically in $\lesssim 1\,{\rm s}$
\citep{alp84a,jah02,sid09,van10}.
Angular momentum stored in the condensate is released erratically
to the crust, causing glitches
\citep{she96},
although the trigger mechanism remains unclear
\citep{war08,mel09,sid10,war11}.

Additionally, differential rotation can occur between spatially
separate regions inside the star.
Examples include inductive and capacitive zones,
invoked to explain the wide range of post-glitch recovery time-scales 
in the same pulsar
\citep{alp96,won01};
regions of strong and weak pinning, predicted by nuclear theory
\citep{jon91,avo07,has11};
and the crust and core
\citep{per05,per06a},
especially if the core is a crystalline color superconductor
\citep{alf08}.
Other condensed astronomical bodies are known or suspected to sustain 
crust-core shear,
notably the Earth, where core super-rotation has been observed
in seismic travel-time data, and the size of the effect and its
variability time-scale are hotly debated
\citep{son96,was11},
and the planets Jupiter and Saturn
\citep{gui99}.

One natural cause of crust-core shear is gravitational stratification.
Oceanographic and other studies of Ekman pumping show that stratification 
limits the volume of fluid that spins up (down) in an impulsively 
accelerated (decelerated) container
\citep{wal69,cla73,stm75,abn96,duc01},
at least over the Ekman time-scale.
\footnote{In this paper, we restrict attention to the spin-down 
problem relevant to an electromagnetically braking neutron star.}
The entire fluid does corotate eventually on a much longer,
dissipative time-scale
\citep{gre63,cla73},
after inertial oscillations die away.
But by then, in a neutron star, 
the crust has decelerated further electromagnetically,
so a crust-core shear always persists.
The shear shapes the gravitational wave signal emitted
during post-glitch relaxation
\citep{van08,ben10}.

In this paper, we apply classic ideas from Solar spin down
\citep{sak71,cla73} to show that the core of a neutron star
rotates systematically faster than the crust in response to the
combined action of stratification-limited Ekman pumping and
electromagnetic braking.
In \S\ref{sec:fas2}, we identify the processes that govern the approach
to a steady state in stratified spin down.
In \S\ref{sec:fas3}, we evaluate the relevant time-scales and estimate
the crust-core lag, finding that it is large early in the star's life
but decreases with age.
In \S\ref{sec:fas4}, we ask whether magnetic crust-core coupling is 
strong enough to erase the crust-core lag and, if so, under what conditions,
a subtle question.
The observational consequences of a persistent lag for precession,
magnetic field evolution, and gravitational radiation 
are outlined briefly in \S\ref{sec:fas5}.
Each consequence deserves to be assessed in detail, 
a challenging task which awaits further work.

\section{Stratified Ekman pumping
 \label{sec:fas2}}
The fluid interior of a neutron star responds to changes in the
angular velocity $\Omega$ of the crust via Ekman pumping.
In an unstratified, Newtonian fluid, Ekman pumping operates as follows:
the Coriolis force drives a secondary circulation in boundary layers 
of thickness $E^{1/2}R$ abutting the crust;
then the boundary flow sucks fluid from the interior, spins it down, 
and recycles it on the Ekman time-scale $t_E = E^{-1/2} \Omega^{-1}$
\citep{gre63}.
The Ekman number,
$E = \nu / (R^2 \Omega)$,
is a function of the kinematic viscosity $\nu$ and stellar radius $R$.
It is the only dimensionless variable in the problem.
In an unstratified, two-component, Hall-Vinen-Bekharevich-Khalatnikov superfluid,
much the same thing happens to the viscous component, 
which in turn drags along the inviscid component via mutual friction
\citep{van10,van11,van12}.

Stratification alters Ekman pumping fundamentally.
In addition to $E$, two new dimensionless variables enter the problem:
the Burger number, 
$S=N^2/\Omega^2$,
which quantifies the relative strength of the buoyancy and Coriolis forces,
and the Prandtl number,
$\sigma=\nu/\kappa$,
which quantifies the relative rate at which momentum and heat are
transported diffusively.
The symbols $N$ and $\kappa$ stand for the 
Brunt-V\"{a}is\"{a}l\"{a} frequency and thermal diffusivity respectively.
We see below that, in a neutron star, $S$ is large
[cf.\ \citet{bay76}], and $\sigma$ is either
small or of order unity, depending on exactly how superfluidity affects $\kappa$.
The interplay and ordering of $E$, $S$, and $\sigma$ lead to
a rich variety of motions
\citep{sak71,cla73,stm75,spe92}.

First and foremost, stratification limits the interior volume influenced by
Ekman pumping to a boundary region, called a buoyancy layer,
of thickness $S^{-1/2}R$
\citep{wal69,cla73}.
In a neutron star, the buoyancy layer is thicker than the unstratified
Ekman layer ($S\ll E^{-1}$),
but it does not suck in, spin down, and recycle interior fluid
to produce uniform rotation. 
Instead, on the Ekman-like time-scale $S^{-1/2} E^{-1/2} \Omega^{-1}$,
the interior organizes itself into a quasi-steady, sheared, azimuthal flow, 
whose angular velocity runs smoothly from its original value deep in the core 
to match the spun-down buoyancy layer at the crust
\citep{sak71,spe92}.
The angular velocity profile is set by the geometry and temperature 
boundary conditions;
in general, Taylor-Proudman columnarity breaks down
\citep{stm75,fri76}.

Secondly, stratification distorts the boundary layer
\citep{wal69}.
In a cylinder, the side wall cannot carry a vertical (i.e.\ falling) mass flux,
so it cannot accept an $O(E^{1/2})$ Ekman inflow.
As a result, fluid is pushed into the corners and thence erupts into the 
interior as a jet, leaving behind a stagnant region
\citep{spe92}.
Eruption also occurs in a sphere, 
with Ekman suction turning into blowing at latitudes $\lesssim 24^{\circ}$
\citep{cla71,fri76}.
The jet exists for $S\sigma >E^{2/3}$, 
which always holds in a neutron star.

Ultimately, of course, the sheared interior flow must evolve 
towards uniform rotation. When does this happen? 
Not until the faster of two dissipative processes takes hold:
viscous diffusion (viscous time-scale $t_\nu = E^{-1} \Omega^{-1}$),
or thermally driven circulation
(Eddington-Sweet time-scale $t_\kappa = \sigma S E^{-1} \Omega^{-1}$,
i.e.\ the thermal diffusion time-scale multiplied by $S$)
\citep{cla73,stm75,hol09}.
If thermal conduction in the neutron superfluid is mainly collisional, 
one has $\sigma \sim 1$
\citep{bai01,urp03}, 
and it is safe to ignore the latter process.
However, in terrestrial superfluids like helium II,
convective counterflow of the small fraction of uncondensed neutrons
boosts the effective thermal conductivity dramatically
\citep{cut87,bai01}.
The possibility that similar physics is at work in a neutron star
is hard to quantify but cannot be dismissed,
so we analyse it too in what follows.
For $\sigma S < 1$, a thermal layer diffuses inwards,
its thickness growing with time $t$ as $(\kappa t)^{1/2}$.
It encounters the inner edge of the buoyancy layer
at $t=\sigma S^{-1} E^{-1} \Omega$,
whereupon the angular velocity decrement imposed by the crust
diffuses inwards within a layer whose thickness grows $\propto t^{1/4}$, 
reaching the center once $t$ equals $t_\kappa$ 
\citep{sak71,cla73,stm75}.

\section{Long-term crust-core shear
 \label{sec:fas3}}
The maximum angular velocity gradient within the star depends on
where the electromagnetic braking time-scale $\Omega/\dot{\Omega}$
fits into the hierarchy of hydrodynamic time-scales.
Over the long term, there are two astrophysically interesting scenarios:
$t_E < \Omega/\dot{\Omega} < {\rm min}(t_\nu,t_\kappa)$
and
${\rm min}(t_\nu,t_\kappa) < \Omega/\dot{\Omega}$.
In the first scenario, early in the star's life,
the angular velocity gradient equals
$(\Omega_0-\Omega)/R$ (if $t_\nu < t_\kappa$)
or
$(\Omega_0-\Omega) (t/t_\kappa)^{1/4} / R$ (if $t_\nu > t_\kappa$),
where $t$ is the star's age,
and $\Omega_0$ is the initial angular velocity;
in other words,
the rotation of the core is a fossil record of its rotation at birth.
In the second scenario, when the star is older, 
the angular velocity gradient equals
$|\dot{\Omega}| \, {\rm min} (t_\nu,t_\kappa) / R$;
all memory of the natal rotation disappears,
and the gradient is proportional to the small amount of spin down
that occurs while the relevant dissipative process unfolds
[see \S6 in \citet{sak71} and \S5 in \citet{cla73}].
\footnote{For $\sigma S \ll 1$, \citet{fri76} found that uniform rotation 
is established on the viscous time-scale $t_\nu$, if the crust
experiences a constant spin-down torque,
contrasting with \citet{sak71},
who found that $t_\kappa$ is the corotation time-scale, 
if the crust decelerates impulsively.
This subtlety is an idiosyncracy of the regime $\sigma S \ll 1$,
which is unlikely to be reached in a neutron star,
even if superfluidity enhances $\kappa$.}

To quantify the shear, we refer to the latest, {\em ab initio} calculations 
of transport coefficients in bulk nuclear matter
\citep{yak99,bai01,sht08}
to evaluate the dimensionless hydrodynamic variables
for a standard neutron star
with mass $\approx 1.4 M_{\odot}$ and $R=10^6\,{\rm cm}$:
\begin{eqnarray}
 E
 & = &
 1.4 \times 10^{-10}
 R_\nu
 \left( \frac{\rho}{\rho_0} \right)
 \left( \frac{T}{10^8\,{\rm K}} \right)^{-5/3}
 \left( \frac{\Omega}{10^2\,{\rm s^{-1}}} \right)^{-1},
\label{eq:fas1}
 \\
 S 
 & = &
 2.9\times 10^2
 \left( \frac{\rho}{\rho_0} \right)^{1/3}
 \left( \frac{\Omega}{10^2\,{\rm s^{-1}}} \right)^{-2},
\label{eq:fas2}
 \\
 \sigma
 & = &
 7.4 \left( \frac{R_\nu}{R_\kappa} \right)
 \left( \frac{\rho}{\rho_0} \right)^{1/3}
 \left( \frac{T}{10^8\,{\rm K}} \right)^{1/3}.
\label{eq:fas3}
\end{eqnarray}
In equations (\ref{eq:fas1})--(\ref{eq:fas3}),
$\rho$ and $T$ denote the density and temperature in the core,
$\rho_0 = 2.8\times 10^{14}\,{\rm g\,cm^{-3}}$ 
is the nuclear saturation density,
and $R_\nu$ and $R_\kappa$ are superfluidity suppression factors
for $\nu$ and $\kappa$ respectively.
\footnote{
In general, $R_\nu$ and $R_\kappa$ are complicated functions of $T$,
with $R_\nu\neq R_\kappa$.
As a rough rule of thumb, one finds
$R_\nu = R_\kappa = 1$ in the absence of superfluidity,
$R_\nu \approx R_\kappa \approx 0.2$ if only the neutrons are
strongly superfluid, and
$R_\nu \approx R_\kappa \approx 0.05$ if both the neutrons and
protons are strongly superfluid.
If convective counterflow in the uncondensed neutrons facilitates
thermal transport, the foregoing approximations still apply to $R_\nu$,
and we have $R_\kappa \gg R_\nu$,
but the value of $R_\kappa/R_\nu$ is not known accurately
\citep{yak99,bai01,sht08}.
}
\footnote{
The core temperature is related to the surface temperature via a
two-zone heat-blanket model \citep{gud82},
and the surface temperature is related to the age $t$ via
`fast' (superfluid) cooling curves (including kaon and pion condensates
and the direct Urca process) calibrated against 11 neutron stars
with measured thermal X-ray emission [see Figure 2 in \citet{pag98}].
\label{foot:fas1}
}
We assume $npe$ matter for simplicity:
the electrons are ideal, degenerate, and ultrarelativistic,
while the nucleons are nonideal and nonrelativistic
\citep{yak99}.
To calculate the thermal conductivity, we assume that 
neutron-neutron and neutron-proton quasiparticle collisions dominate
\citep{bai01}
(but set the dressed masses equal to the bare masses for simplicity),
neglect electron conduction,
neglect superfluidity in the scattering frequencies (but not elsewhere),
and calibrate the result against Figure 1 in the latter reference. 
To calculate the shear viscosity, we assume that
electron-electron collisions dominate,
except at $T \sim 10^7\,{\rm K}$ where neutrons also contribute,
we neglect muons (consistent with the $npe$ assumption),
and we include the important effect of
Landau damping mediated by weakly screened transverse plasmons
\citep{sht08}.
Calibrating against Figure 4 in the latter reference, we see that
Landau damping lowers $\nu$ by nearly an order of magnitude with respect to
the traditional value
\citep{cut87},
and $\nu$ scales as $T^{-5/3}$ instead of $T^{-2}$, 
which is astrophysically significant because it leaves a $T$ dependence in $\sigma$.
The Brunt-V\"{a}is\"{a}l\"{a} frequency, arising from departures from
beta equilibrium, is proportional to the square root of the 
electron-neutron ratio and is taken from \citet{rei92}.

\begin{figure}
\begin{center}
\includegraphics[width = 15cm,angle=0]{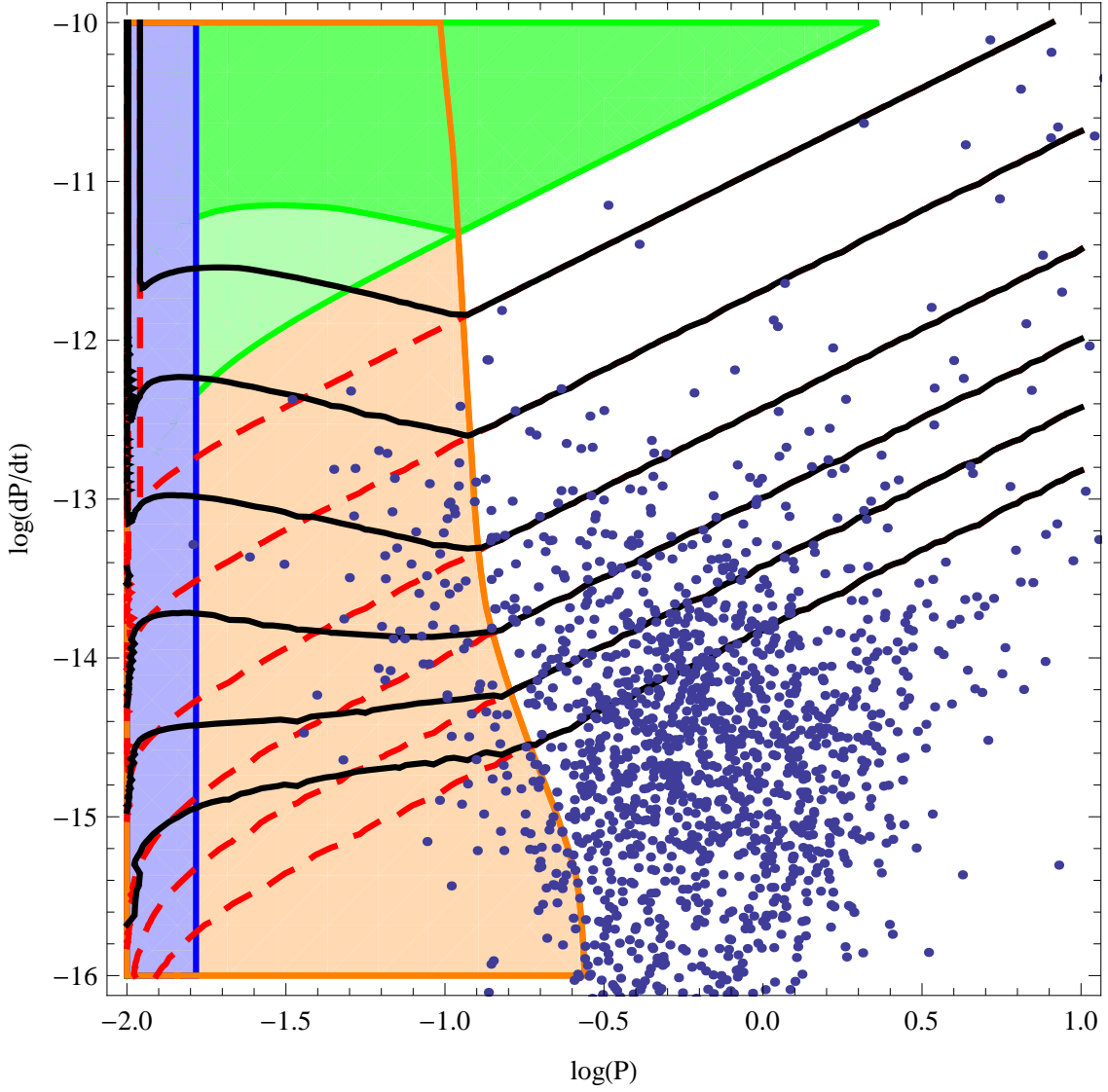}
\end{center}
\caption{
Contour plot of Rossby number as a function of spin period $P$
(horizontal log axis, in ${\rm s}$)
and period derivative $\dot{P}$
(vertical log axis, in ${\rm s\,s^{-1}}$),
running from ${\rm Ro}=0.1$ (top contour) to $10^{-6}$ (bottom contour) 
in steps of 1 dex.
Red, dashed curves:
strong neutron and proton superfluidity, 
collisional thermal conduction
($R_\nu = R_\kappa = 0.05$).
Black, unbroken curves:
counterflow-enhanced thermal conduction
($R_\nu = 0.05$, $R_\kappa = 5\times 10^2$).
Blue dots mark known radio pulsars.
The colored regions cover the following regimes.
Blue: 
$S<20$, 
stratification too weak to impede Ekman pumping, ${\rm Ro}$ negligible.
Heavy green:
fossil stage,
$t < t_\nu < t_\kappa$.
Light green:
fossil stage,
$t < t_\nu < t_\kappa$ ($t < t_\kappa < t_\nu$)
for $R_\kappa = 0.05$ ($5\times 10^2$).
Orange:
$t>{\rm min}(t_\nu,t_\kappa) = t_\kappa$,
natal rotation erased,
${\rm Ro}$ decreases monotonically with age,
thermally driven circulation pre-empts viscous diffusion.
White:
$t>{\rm min}(t_\nu,t_\kappa) = t_\nu$,
viscous diffusion pre-empts thermally driven circulation.
Other parameters:
$\rho=\rho_0$, $\Omega_0=200\pi \, {\rm s^{-1}}$ (fiducial),
$T$ from \citet{pag98} (see footnote \ref{foot:fas1}).
}
\label{fig:fas1}
\end{figure}

Figure \ref{fig:fas1} illustrates how much core super-rotation
to expect in neutron stars of different ages. 
The results are expressed in terms of the Rossby number,
${\rm Ro} = \Delta\Omega/\Omega$,
where $\Delta\Omega$ is the angular velocity of the core relative to
the crust.
One obtains
${\rm Ro} = \Omega_0 / \Omega - 1$ and 
${\rm Ro} = |\dot{\Omega}| \, {\rm min} (t_\nu,t_\kappa) / \Omega$ 
respectively in the two scenarios described at the start of \S\ref{sec:fas3}.
Contours of ${\rm Ro}$ are graphed
on the plane spanned by the rotation period $P$
and its time derivative $\dot{P}$.
Known radio pulsars are overplotted as blue dots.
The blue band at the left of the diagram marks the excluded region 
$S\lesssim 1$,
where stratification is weak, buoyancy does not impede Ekman pumping,
there is no fossil rotation, the long-term shear 
${\rm Ro} \approx t_E/t \lesssim 10^{-6}$ is too small to be of interest,
and the analysis in this paper does not apply.

Consider first the example $R_\nu = R_\kappa = 0.05$ (red, dashed contours),
which corresponds to strong superfluidity in the neutrons and protons,
without counterflow-enhanced thermal conduction.
In this regime, we find $t_\nu < t_\kappa$; 
viscous diffusion pre-empts thermally driven circulation throughout
the displayed subset of the $P$-$\dot{P}$ plane,
and the orange shading in the figure (explained in the next paragraph)
is irrelevant.
In most of the figure, below and to the right of the green shading,
we also have $t>t_\nu$ and hence ${\rm Ro} = t_\nu/(2t)$.
The red contours run from ${\rm Ro} = 10^{-1}$ to $10^{-6}$, spaced by one dex.
Numerous objects with spin-down ages $\lesssim 10^4\,{\rm yr}$
are predicted to have relatively high Rossby numbers.
In particular, there are three objects with ${\rm Ro} \geq 10^{-1}$
(the young pulsar PSR J1846$-$0258 in the supernova remnant Kesteven 75,
and the magnetars PSR J1550$-$5418 and PSR J1907$+$0919),
and 18 with $10^{-2} \leq {\rm Ro} \leq 10^{-1}$
(including three with ${\rm Ro}\approx 0.1$).
PSR J1846$-$0258, which is only $0.7\,{\rm kyr}$ old,
and whose braking index has been measured by absolute pulse numbering,
is an especially good rotation fossil candidate
\citep{got00,liv06}.
In the top left corner of the figure, outlined by the green triangle,
where young pulsars with $t\lesssim 10^3\,{\rm yr}$
and $T\gtrsim 10^8\,{\rm K}$ reside,
we find $t < t_\nu$ and hence predict ${\rm Ro} \gtrsim 1$,
i.e.\ the core retains much of its natal angular velocity.
No radio pulsars have been discovered in the green triangle to date,
but such objects, whether young like PSR J1846$-$0258 or magnetar-like, 
are easily accommodated within our current picture of neutron star evolution.

Next consider the example $R_\nu = 0.05$, $R_\kappa = 5\times 10^2$ 
(black, unbroken contours),
which models roughly the counterflow effect.
The orange shading now becomes relevant; 
it divides the $P$-$\dot{P}$ plane into two parts. 
In the white region to the right of the orange curve,
we find $t_\nu < t_\kappa$, and nothing changes from the previous example:
viscous diffusion dominates, and the black ${\rm Ro}$ contours overlap
the red ones.
To the left of the orange curve, however, the opposite is true:
we find $t_\kappa < t_\nu$, 
and thermally induced circulation occurs faster than viscous diffusion.
Below the irregularly shaped, heavy green region,
we also have 
$t>
 t_\kappa
 = 9.7(R_\kappa/5\times 10^2)^{-1}
 (\rho/\rho_0)^{-1/3} 
 (T/10^8\,{\rm K})^2 
 (\Omega/10^2\,{\rm s^{-1}})^{-2}
 \, {\rm yr}$
and hence ${\rm Ro} = t_\kappa/(2t)$.
That is, when $R_\kappa$ is large, 
the green region to the right of the orange curve ($t_\nu < t_\kappa$)
is a promising hunting ground for rotation fossils with 
${\rm Ro} \geq 1$;
we do not expect to find many to the left of the orange curve.
Likewise, when $R_\kappa$ is large, it is harder to find objects 
with ${\rm Ro}$ less than unity but still relatively large;
as the black contours tilt upwards in the orange region, 
one needs a younger, faster pulsar to achieve a given ${\rm Ro}$
than when $R_\kappa$ is small,
yet young pulsars spin down quickly.
A meaningful fraction of the pulsar population lies within the orange region.
No known objects are predicted to have ${\rm Ro}\geq 0.1$,
three have $10^{-2} \leq {\rm Ro} \leq 10^{-1}$
[the gamma-ray pulsar PSR J1023$-$5746, 
discovered in a blind Fermi Large Area Telescope search
\citep{saz10},
PSR J0540$-$6919 in the Large Magellanic Cloud, and the Crab],
and 16 have $10^{-3} \leq {\rm Ro} \leq 10^{-2}$

Figure \ref{fig:fas1} also maps the history of differential rotation
in individual objects. As a pulsar brakes electromagnetically, 
it moves from the top left to the bottom right of the figure along
a diagonal spin-down trajectory, $P\dot{P} = {\rm constant}$.
If the trajectory intersects the green triangle,
the early part of the star's life is spent as a rotation fossil.
\footnote{
For the minority of neutron stars born
in the blue region, where stratification is weak, Ekman pumping 
erases the natal rotation on the fast time-scale $t_E$.
}
When the star crosses from the green region into the orange
or unshaded regions, ${\rm Ro}$ drops suddenly
on the time-scale
${\rm min}(t_\nu,t_\kappa)$,
as dissipative processes take over.
Subsequently, 
the crust-core shear decreases monotonically as the star brakes and cools, 
with $\dot{\Omega}/\Omega \propto t^{-1}$ and hence
${\rm Ro} \propto t^{-1} T^{5/3}$ or
${\rm Ro} \propto T^{2}$
in the unshaded or orange regions respectively.
\footnote{
The transition between the two scalings occurs smoothly at the wavy
orange curve in Figure \ref{fig:fas1}.
}
If the trajectory starts outside the green triangle,
the latter behavior characterizes the star's whole life:
there is no fossil stage.
A wide range of terminal ${\rm Ro}$ values are possible,
when the trajectory intersects the pair-cascade death line
\citep{hib01a}.

\section{Magnetic coupling
 \label{sec:fas4}}
How does the stellar magnetic field alter the spin-down scenario above? 
This question turns out to be subtle.
The physics depends sensitively on the internal magnetic topology and 
the multi-component nature of the superfluid interior,
both of which remain highly uncertain. A proper understanding also requires 
a theory of Ekman pumping in a stratified, magnetized fluid in a sphere, 
which does not exist yet in the literature even for uniform magnetization,
let alone for a realistic (e.g. linked poloidal-toroidal) field geometry
or for a superconductor.
In this section, therefore, we restrict ourselves to canvassing some of the 
main issues and presenting associated order-of-magnitude scalings, 
noting that large-scale numerical simulations will ultimately be required 
to settle the matter decisively.

Magnetic coupling in its simplest guise 
(i.e.\ in an unstratified, uniformly magnetized star) speeds up angular
momentum transport dramatically, enforcing corotation between the
crust and core over $\sim 10^2\,{\rm s}$ for typical magnetic field strengths
and spins.
The process occurs in two stages. Firstly, the charged component of the
superfluid is accelerated by Lorentz forces, mediated not by Alfv\'{e}n waves
but rather by rotation-dominated, hydromagnetic-inertial waves,
if the star is not superconducting 
\citep{eas79a},
or by cyclotron-vortex waves, whose restoring force comes from
the vortex tension, if the star is superconducting
\citep{men98}.
Secondly, the charged fluid drags along the neutron condensate
via mutual friction, as charges scatter off neutron vortices
magnetized by the Fermi liquid interaction (entrainment)
\citep{alp84a}.
Overall, for uniform magnetization, the coupling time-scale is given by
$t_B = 2(V_A/R \Omega)^{-2/3}\Omega^{-1}
= 6.6
 (\rho/\rho_0)^{1/3}
 (\Omega/10^2\,{\rm s^{-1}})^{-1/3}
 (B/10^{12}\,{\rm G})^{-2/3} 
 \, {\rm s}$ 
in a non-superconducting star \citep{eas79a},
where $V_A = (B^2/4\pi\rho)^{1/2}$ is the Alfv\'{e}n speed,
and $B$ denotes the magnetic field strength.
In a type II superconductor, $B^2$ is replaced by $BH_{c1}$
in the formula for $t_B$, where $H_{{\rm c}1} \sim 10^{15}\,{\rm G}$ 
is the lower critical field, strengthening the coupling further.

In a stratified star, the magnetized Ekman problem remains unsolved.
Broadly, though, stratification (and compressibility) inhibit
crust-core coupling, just as they do in an unmagnetized star
\citep{abn96}.
To estimate the stratified magnetic coupling time-scale, 
we follow the pioneering order-of-magnitude analysis in \S{6} of \citet{men98},
which is arguably somewhat representative even though it applies to the idealized case 
of uniform magnetization.
\citet{men98} showed that the $t_B$ formula in the
previous paragraph applies only to the uppermost, buoyancy-limited
boundary layer adjacent to the crust. One then imagines a stack
of $\approx R\Omega/V_A$ buoyancy-limited layers 
reaching from the crust to the core,
which are activated sequentially and combine to give a revised time-scale,
$t_B = 2(V_A/R \Omega)^{-5/3}\Omega^{-1}
= 3.9\times 10^4
 (\rho/\rho_0)^{5/6}
 (\Omega/10^2\,{\rm s^{-1}})^{2/3}
 (B/10^{12}\,{\rm G})^{-5/3} 
 \, {\rm s}$.
Hence it is possible to have $t_B > t_\nu$ ---
and hence maintain core super-rotation on the viscous time-scale,
as in \S\ref{sec:fas2} and \S\ref{sec:fas3} ---
in objects with moderately magnetized interiors and/or fast spins,
satisfying
$(B/10^{12}\,{\rm G})^{5/3} 
 (\Omega/10^2\,{\rm s^{-1}})^{-2/3}
 <
 5.5\times 10^{-4} R_\nu (\rho/\rho_0)^{11/6}
 (T/10^8\,{\rm K})^{-5/3}$.
Such objects may be found squarely within the main pulsar population
in Figure \ref{fig:fas1}
(e.g. $B\lesssim 10^{10}\,{\rm G}$)
or among the recycled millisecond pulsars, 
provided that the internal fields in the latter are actually reduced 
rather than just buried,
cf.\
\citet{kon97} and \citet{pay04}.
The above conclusions are consistent with related work on boundary-layer
damping of r-modes.
In unmagnetized r-modes, it is found that compressibility approximately
halves the damping rate, while stratification has almost no effect, 
because the radial fluid velocity component in an r-mode is tiny
\citep{gla06},
unlike in the spin-down problem considered here, where the radial
and angular velocity components are of the same order and stratification
matters
\citep{abn96}.
In magnetized r-modes, analyzed for the idealized, unstratified problem
of a radial magnetic field with uniform or cosine-latitude magnitude,
Lorentz forces shorten the damping time-scale,
offset partially by the superfluid mutual friction,
because damping occurs in a boundary layer abutting the crust,
where buoyancy is negligible
\citep{men01,kin03},
unlike in core spin down.

In a stratified star {\em with a realistic internal magnetic geometry},
things are more uncertain.
Arguably, however, the overall impact of a nontrivial geometry
is to lengthen the magnetic coupling time-scale further.
One can see this qualitatively by considering the dispersion relations
of hydromagnetic waves in a stratified slab, where the stratification axis
${\bf n}$ is inclined obliquely to both the background magnetic field
${\bf B}_0$ and the direction ${\bf k}$ of wave propagation.
A thorough treatment of this problem in plane-parallel geometry
(without rotation, unfortunately)
is given for various wave modes in Chapter 7 of 
\citet{goe04}.
The details fall outside the scope of this paper, but generically
one discovers that the phase speed is typically lower than the
buoyancy-limited Alfv\'{e}n speed in a uniform 
magnetic field and is much lower for certain orientations.
For example, when ${\bf n}$ is perpendicular to ${\bf k}$ and ${\bf B}_0$,
and the wave propagates obliquely with respect to ${\bf B}_0$,
the phase speeds of the Alfv\'{e}n and slow modes
(which are favored because they displace fluid horizontally,
unlike the fast mode) tend to zero in the small-beta
(pressure-dominated) regime relevant to neutron stars, 
as illustrated in Figure 7.10 of \citet{goe04}.
In the uniform-${\bf B}_0$ calculations attempted so far in the
neutron star literature, such orientations are avoided
by construction but they are inevitable in realistic geometries.
For example, in a linked poloidal-toroidal structure
\citep{bra06a,lan12}, where field loops close inside the star,
many internal surfaces exist where unfavorable orientations
of ${\bf n}$, ${\bf k}$, and ${\bf B}_0$ are achieved,
and buoyancy-modified hydromagnetic waves slow down dramatically,
lengthening the crust-core magnetic coupling time. In a tangled field,
this is even more likely to occur 
\citep{bra06a}.
\citet{goe04} emphasized in {\S}7.3.3 of their book that these subtle effects
emerge fully when oblique propagation 
($0 < | {\bf \hat k} \cdot {\bf \hat B}_0 | < 1$) is considered, 
and that they are expected to be even more prevalent when ${\bf n}$
is inclined obliquely to ${\bf B}_0$ as well, the relevant scenario
in general in a neutron star [and outside the scope of \citet{goe04}].
There is no substitute for a self-consistent, three-dimensional
numerical simulation to quantify these difficult matters. 
The prospects are fair that suitable codes and computational platforms 
will become available shortly that are equal to the task.

In a realistic magnetic geometry, another crucial complication arises:
many field topologies are incapable of enforcing crust-core corotation at all,
no matter how fast the magnetic coupling.
Quoting from \citet{eas79b}, 
who demonstrated this counter-intuitive property analytically,
``the field must have a topology such that the equation for the
perturbation ${\bf B}'$ in the field induced by the slowing down
has nonsingular, single-valued solutions''.
At first blush, this sounds like a side-issue of purely academic interest,
but nothing could be further from the truth.
There exist at least two important topologies 
(which are popular in the literature, incidentally)
that violate the above condition and cannot enforce corotation:
(1) axisymmetric, poloidal fields with loops closing inside the star;
and (2) axisymmetric, toroidal fields.
To see why, consider the momentum equation in superfluid 
magnetohydrodynamics, viz.\ equation (1) in \citet{eas79b}.
Several of the forces are not conservative in general,
e.g.\ the Lorentz force and the Fermi liquid interaction.
\footnote{
The entrainment force looks superficially like it is constructed
out of gradients of the internal energy and proton effective mass,
but the prefactors multiplying these gradients, like the proton
number density and pressure, are functions of position even in equilibrium
due to compositional stratification;
see equation (2) in \citet{eas79b}.
}
The nonconservative forces cannot be balanced everywhere
by pressure gradients and buoyancy, so ${\bf B}'$ tries to adjust
until a balance is achieved. But often there is no way to do so.
For example, if ${\bf B}_0$ is poloidal and axisymmetric,
then ${\bf B}'$ is toroidal and the toroidal ($\varphi$-)
component of the momentum equation reduces to
$(4\pi)^{-1} ({\bf B}_0\cdot\nabla)(\varpi B_\varphi')
= F({\bf q},\varpi)$
in equilibrium,
where $\varpi$ denotes the cylindrical radius, 
and the right-hand side is a complicated nonconservative function 
$F(\dots)$ of unperturbed variables ${\bf q}$
like the neutron and proton mass densities,
viz.\ equation (6) in \citet{eas79b}.
Integrating along a poloidal field line that closes inside the star,
one reaches the unphysical conclusion that $\varpi B_\varphi'$ is multi-valued.
It is tempting to presume that the plasma can always find some way
to adjust slightly and quasi-statically near a sensible topology
to fix the problem, 
``so that the relevant integrals along all closed loops give zero''
\citep{eas79b}.
But \citet{eas79b} proved directly from the hydromagnetic equations of motion 
that the adjustment fails to achieve balance except in very special situations,
because $F({\bf q},\varpi)$ is a function of variables
that depend only on the {\em unperturbed} state of the star
(density, temperature, ${\bf B}_0$, and the crust angular velocity,
which do not respond as ${\bf B}'$ adjusts)
as well as the angular velocity of the superfluid vortex array, 
which is governed by a superfluid equation of motion 
independent of ${\bf B}'$.
These important points are not appreciated widely in the literature 
and deserve careful study in future. 
For example, is the set of fields ${\bf B}_0$
that give unphysical solutions for ${\bf B}'$ ``large'' or ``small''?
\citet{eas79b} expressed a hope that numerical simulations would
answer this question, but the problem is formidable, 
and the relevant simulations have not yet been done.

Magnetic flux is concentrated into quantized flux tubes in a type II superconductor,
modifying the magnetohydrodynamic equations of motion
\citep{gla11}.
The implications for core super-rotation remain highly uncertain.
One outcome is to replace $B^2$ by $BH_{{\rm c}1}$ in the classic formula
for the magnetic coupling time-scale, as described at the start
of this section, but even this result has not yet been confirmed in
realistic geometries and in the presence of stratification,
where other length-scales and directional effects may enter the problem.
Another outcome is that the flux tubes are expected to interact with
the superfluid vortices in the neutron condensate, 
which are magnetized by entrainment.
Opinions remain divided as to whether the interaction is strong,
locking together the charged and neutral components
\citep{rud98},
or weak, 
as in the recent ``snowplow'' model of glitches
\citep{has12,sev12},
if the core is a type I superconductor
\citep{jon06},
contains hyperons
\citep{bab09},
or is the seat of vortex instabilities
\citep{per06a,and07,lin12}.
We do not attempt to adjudicate these difficult matters in this paper.
Efforts are currently under way to calculate analytically 
the Ekman coupling time-scale
in a type II superconductor including some of the above physics, 
albeit in an idealized plane-parallel geometry and a uniform magnetic field.
\footnote{
K.\ Glampedakis, private communication.
}

Whether strong or weak, magnetic coupling does not engage the uncondensed neutrons, 
which are viscous, participate in Ekman pumping, carry inertia,
and therefore play a role in maintaining core super-rotation
[and in explaining radio timing data of pulsar glitch recoveries;
see \citet{van10}].
It is sometimes assumed that the uncondensed neutrons carry a tiny
fraction of the star's inertia, because their density is proportional
to $\exp(-\Delta/k_{\rm B}T)$ in the ideal Bardeen-Cooper-Schrieffer theory,
and $k_{\rm B}T$ is much smaller than the superfluid gap $\Delta$ for most 
neutron pairing schemes
\citep{yak99}.
However, the ideal exponential gap formula tells only part of the story.
Applied to superfluid helium, 
it predicts zero uncondensed atoms as $T\rightarrow 0$,
whereas the experimentally measured fraction is $\approx 14\%$
due to van der Waals forces.
It is likely that the nuclear color force is nonideal in the same way,
as the quarks and gluons inside nucleons are polarized by random motions
of nearby nucleons at high densities
\citep{web05}.
The effect has not been quantified yet from first principles,
as it requires a hard calculation in quantum chromodynamics, 
but there are good indications that
a van der Waals equation of state is consistent with measurements of
the nuclear liquid-gas transition 
\citep{tor10}
and also with scattering experiments
\citep{kai97}.
If the uncondensed fraction is substantial, it helps maintain 
residual core super-rotation via slow, unmagnetized,
stratification-limited Ekman pumping, even when the other stellar components
couple magnetically on a fast time-scale.
\footnote{
Recently an analytic theory of unmagnetized superfluid spin down
has been developed in arbitrary geometry which self-consistently incorporates 
Ekman pumping, mutual friction, and the back reaction of the fluid on its container.
The theory reproduces the results of classic laboratory experiments on liquid helium 
\citep{tsa80}
to $0.5\%$ accuracy with zero free parameters
\citep{van11},
yields excellent fits to Vela's quasi-exponential glitch recoveries,
and predicts without adjustment the ``overshoot'' observed in the Crab
\citep{van10},
all with zero magnetic coupling.
Of course, nonzero magnetic coupling is also consistent with the glitch data,
provided that vortex pinning in the inner crust is invoked to retard
the response;
one is then led to postulate multiple pinning zones to match the
multiple recovery time-scales observed in individual objects
\citep{alp84a}.
}

\section{Observational signatures
 \label{sec:fas5}}
How might one observe a crust-core shear of the magnitude 
in Figure \ref{fig:fas1}?
The question is too multi-faceted to answer properly all at once,
but we tender a few suggestions.
Taking inspiration from the Earth, where seismic data carry the imprint
of the super-rotating core
\citep{son96,zha05,was11},
one option is to look for rotational splitting of the
quasi-periodic oscillations observed in the X-ray light curve 
in the immediate aftermath of a magnetar giant flare
\citep{str06}.
At present, there is a lively debate as to whether magnetar oscillations
arise from shear modes in the crust, magneto-inertial modes in the core,
a combination of the two, or something else entirely
\citep{lan10}.
Theoretical treatments invariably start by perturbing a uniformly rotating
background structure, and the rotational splitting is assumed to be small,
as magnetars are slow rotators.
However, in objects where at least some of the natal rotation of the core 
is preserved as in Figure \ref{fig:fas1},
one might hope to see greater rotational splitting.

Another possible signature is precession. 
A body composed of a rigid crust and liquid mantle precesses
in a superposition of two modes (four if it also contains a
solid inner core) excited by differential rotation,
namely a Chandler wobble and free-core nutation
\citep{kit08a};
think of an egg, whose yolk rotates differentially.
The mode frequencies are related to the inertia of the components 
and the strength of crust-core coupling
via magneto-inertial
\citep{bay76}
or elasto-inertial
\citep{bas99}
waves.
The model can be matched to the harmonics observed in PSR B1828$-$11 
\citep{sta00,kit08a}
\footnote{
The oscillatory timing residuals and pulse profile changes 
in PSR B1828$-$11 are now interpreted in terms of
switching between two magnetospheric states
\citep{lyn10}.
}
and to similar periodic phenomena ranging from months to years 
in other objects
\citep{kit08b}
to infer the crust-core shear in principle.
A long-period crust-mantle wobble may also be excited 
by a rotational glitch, 
when a portion of the crust-core shear is impulsively reset
\citep{yua10}.

If verified observationally, core super-rotation enriches neutron star 
astrophysics
in two important ways.
First, it implies
ongoing, self-limiting, toroidal magnetic field generation.
One might argue that magnetic stresses should react
to erase the shear on the time-scale 
${\rm Ro}^{-1} \Omega^{-1}$,
but the Sun provides a clear counter-example.
Many resistive and other processes intervene to pinch off
toroidal magnetic flux tubes before the back-reaction becomes too strong,
e.g.\ the current-induced Tayler instability, which is especially
active for high magnetic Prandtl numbers
(and is modified by the Hall effect in magnetars)
\citep{rud09},
or magnetic self-organization into a torus-like configuration,
where $\Omega$ is constant along field lines,
winding stops, and the magnetorotational instability is avoided
\citep{due06}.
In numerical simulations of protoneutron stars,
it is standard to find that the magnetic back-reaction saturates 
at $\lesssim 1\% $ of the mechanical shear stress
\citep{coo03,due06}.
Whether core super-rotation also drives a turbulent, self-sustaining,
hydromagnetic dynamo in the face of strong stratification is less clear.
Shear-driven turbulence in a stratified spherical shell automatically
supplies the non-zero helicity which a dynamo requires
\citep{gel11},
and the magnetic Prandtl number easily exceeds the dynamo threshold
$\sim 10^2 E^{3/4}$ in a neutron star.
However, the interplay between $S$ and $\sigma$ is not understood fully;
recent work suggests that the dynamo depends weakly on $S$
and generates a predominantly dipolar field except for ${\rm Ro}\sim 1$,
but the jury is still out
\citep{sim11}.
\footnote{
\citet{bra06b} demonstrated simultaneously the twin phenomena of
self-limiting toroidal field generation and a self-sustaining
dynamo loop within a unified, self-consistent simulation of a differentially
rotating, nonconvective star with and without stable stratification.
Tayler and magnetorotational instabilities disrupt the toroidal field
well before it grows to equipartition, provided that the magnetic diffusion
length-scale is shorter than the buoyancy length-scale (always true
in a neutron star). 
Counter-intuitively, in the stratified system,
a self-sustaining dynamo establishes itself more easily, 
the toroidal field builds up to saturation more slowly, 
and at saturation the toroidal field fluctuates less.
In a statistically steady state, the magnetic back-reaction saturates
at $\approx 5\%$ of the average magnetic pressure, 
much less than the mechanical stress driving the differential rotation.
}
Two potentially observable manifestations
are suggested by terrestrial analogies:
geomagnetic jerks,
wherein small changes in the length of the day are communicated to
the magnetic field via a Hartmann current system connecting the
Ekman/buoyancy layer to the interior
\citep{dav94,dav97},
and longitudinal drift of equatorial geomagnetic spots,
driven by zonal flows over $\sim 10^7$ rotation periods
\citep{sta08,sim11}.
\footnote{
\citet{lop71} argued that magnetic and buoyancy forces can conspire
to mimic unstratified Ekman pumping under special circumstances,
clouding the picture further.
\citet{jie95} agreed but showed that the quasi-steady, sheared flow
in the interior persists much longer than $t_E$.
}
Spot or pole drift occurs over roughly the same time-scale
as the magnetospheric switching observed in objects like PSR B1828$-$11
\citep{lyn10}.

Secondly, core super-rotation opens up new channels of gravitational-wave emission
\citep{van08,ben10}. 
The free energy in the sheared flow can excite
nonaxisymmetric instabilities,
e.g.\ the herringbone, Taylor-G\"{o}rtler, and Stewartson layer modes 
familiar from Newtonian and superfluid spherical Couette flow
\citep{nak02,hol03,per06b,per09}
and their stratified counterparts
\citep{smi05,mun10}.
\footnote{
For example, \citet{per09} found ${\rm Ro} \gtrsim 4 E^{0.4}$ 
from numerical simulations for superfluid Stewartson layer instabilities,
which is satisfied throughout much of Figure \ref{fig:fas1}.
}
\footnote{
Nonaxisymmetric flows excited in the core at birth should also persist
up to the dissipation time-scale ${\rm min}(t_\nu,t_\kappa)$
\citep{ben10}.
}
The theoretical machinery exists to treat such instabilities in a neutron star context,
including stratification, compressibility, magnetic fields, and a differentially rotating 
background in general
\citep{men01,gla06},
but it has only been applied to r-modes on a uniformly rotating background
at the time of writing.
\footnote{
Stratification is unimportant for r-mode damping,
because r-mode circulation is nearly nonradial, unlike spin down.
Compressibility roughly doubles the r-mode damping time
\citep{gla06}.
}
Core super-rotation also complicates gravitational-wave detection:
the gravitational-wave phase may drift relative to the rotational phase inferred
from radio pulse timing.
Finally, as the Reynolds number $E^{-1}$ is large,
the interior flow is likely to be turbulent, even when $S$ is large
\citep{gre70,mel07,mel10}.
Stratified turbulence collapses down to two dimensions (i.e.\ spherical shells) 
only when the activity parameter
${\rm Ro}^3 S^{-1} E^{-1}$ drops below $\approx 7$,
which condition is avoided for ${\rm Ro} \gtrsim 10^{-2}$,
i.e.\ throughout much of Figure \ref{fig:fas1}
[see \citet{iid09} and {\S}2 in \citet{mel10}].
Stochastic gravitational radiation from hydrodynamic turbulence imposes
a fundamental noise floor on gravitational-wave observations of fossil rotators
with ${\rm Ro} \gtrsim 1$ 
even with the current generation of long-baseline interferometers
\citep{mel10}.

In summary, stratification limits Ekman pumping to a thin buoyancy layer
in a neutron star, leaving a super-rotating core as the crust brakes
electromagnetically.
The crust-core shear grows, until dissipation
(viscous or Eddington-Sweet) kicks in, then diminishes.
Of the radio pulsars discovered to date, a handful of young objects
(some Crab-like, some magnetar-like)
are predicted to have unexpectedly high Rossby numbers
${\rm Ro} \gtrsim 10^{-1}$,
i.e.\ their cores are a fossil record of their angular velocities at birth.
It is realistic to think that more candidate fossils will be
discovered by future surveys.
Moreover, there are 24 objects with ${\rm Ro} \geq 10^{-2}$ already known,
some dominated by viscous diffusion, others by thermally driven circulation,
which should be examined as a group for common features.
If core super-rotation is confirmed, it will prompt a review of 
traditional ideas concerning how magnetic fields and gravitational waves
are generated by neutron stars.

\acknowledgments
This research was supported by a CSIRO-University of Melbourne Collaborative Grant.
I thank R. Hoey, K. Poon, and C. Peralta for discussions on
numerical simulations of stratified flows,
M. Bennett and C. van Eysden for discussions on stratified Ekman pumping
in gravitational-wave applications and glitch recovery,
and D. Yakovlev and R. Sawyer for references concerning
thermodynamic and transport coefficients in bulk nuclear matter.

\bibliographystyle{mn2e}
\bibliography{stratified}

\end{document}